\documentclass[aps,twocolumn,showpacs,preprintnumbers]{revtex4}

\newcommand{\nc}{\newcommand}
\nc{\rnc}{\renewcommand}
\nc{\beq}{\begin{equation}}
\nc{\eeq}{\end{equation}}
\nc{\bea}{\begin{eqnarray}}
\nc{\eea}{\end{eqnarray}}
\nc{\ba}{\begin{array}}
\nc{\ea}{\end{array}}
\nc{\nn}{\nonumber}
\nc{\bpi}{\begin{picture}}
\nc{\epi}{\end{picture}}
\nc{\scs}{\scriptstyle}
\nc{\unit}{{\mbox{\boldmath\large $1$}}}
\nc{\p}{\partial}
\nc{\ua}{\uparrow}
\nc{\da}{\downarrow}
\nc{\uada}{{\uparrow\downarrow}}

\nc{\al}{\alpha}
\nc{\be}{\beta}
\nc{\ga}{\gamma}
\nc{\de}{\delta}
\nc{\la}{\lambda}
\nc{\si}{\sigma}
\nc{\Ga}{{\sf\Gamma}}
\nc{\La}{\Lambda}

\nc{\abar}{\bar{a}}
\nc{\J}{{\sf J}}
\nc{\T}{{\sf T}}
\rnc{\P}{{\sf P}}
\nc{\R}{{\sf R}}
\rnc{\S}{{\sf S}}
\nc{\U}{{\sf U}}
\nc{\V}{{\sf V}}
\nc{\Va}{{\sf V}_a}
\nc{\Vap}{{\sf V}_a'}
\nc{\Vb}{{\sf V}_b}
\nc{\Vc}{{\sf V}_c}
\nc{\Vd}{{\sf V}_d}
\nc{\X}{{\sf X}}
\nc{\Y}{{\sf Y}}
\nc{\Z}{{\sf Z}}

\nc{\cd}[2]{\scs\{\raisebox{-.46ex}{\rlap{\tiny{#2}}}
  \raisebox{.45ex}{\tiny{#1}}\scs\}}
\nc{\zm}{\ba{cc}0&0\\0&0\ea}
\nc{\zr}{\ba{cc}0&0\ea}
\nc{\zc}{\ba{c}0\\0\ea}

\begin{document}

\bibliographystyle{apsrev}

\title{Simplifying Kaufman's Solution of the Two-Dimensional Ising Model}
\author{Boris Kastening}
\email[Email address: ]{ka@physik.fu-berlin.de}
\affiliation{Institut f\"ur Theoretische Physik\\
Freie Universit\"at Berlin\\
Arnimallee 14\\
D-14195 Berlin\\
Germany}

\date{2 August 2001}

\begin{abstract}
We considerably simplify Kaufman's solution of the two-dimensional Ising
model by introducing two commuting representations of the complex rotation
group SO$(2n,C)$.
All eigenvalues of the transfer matrix and therefore the partition
function are found in a straightforward way.
\end{abstract}

\pacs{05.50.+q, 68.35.Rh, 05.70.Ce, 64.60.Cn.
Published in Phys.~Rev.~E64, 066106 (2001).}

\maketitle

Since Onsager's solution \cite{onsager} in the transfer matrix approach
\cite{kw} of the two-dimensional Ising model \cite{li} with vanishing
magnetic field and its subsequent simplification by Kaufman \cite{kaufman},
there have been a number of related as well as alternative solutions, see
e.g.\ Baxter's book on exactly solved models in statistical mechanics
\cite{baxterbook} and references therein.
Among the transfer matrix solutions are the ones by Schultz, Mattis
and Lieb, by Thompson, by Baxter and by Stephen and Mittag \cite{tmf}.
Nevertheless, the author of this work feels that there is still room for
a nice and straightforward solution.
A completely self-contained and detailed account of this work may be
found in \cite{bo}.

We study the two-dimensional Ising model with zero magnetic field on a
square lattice with $m$ rows and $n$ columns subject to toroidal
boundary conditions.
The transfer matrix is expressed in terms of the generators of two
commuting representations of the complex rotation group SO$(2n,C)$.
These representations naturally arise from projected bilinears of
$2^n\times2^n$ spin matrices.
Conservatively speaking, we reduce Kaufman's approach to its essential
steps, avoiding in particular the doubling of the number of eigenvalues
of the transfer matrix and subsequent rather involved arguments for the
choice of the correct ones.
Additionally, there is no need to investigate the transformation properties
of the spin matrices.

Our notation is in the spirit of \cite{huang}, with sans serif capitals
reserved for $2^n\times2^n$ matrices.
The structure of this work is as follows:
After defining the model and its transfer matrix $\T$, we express
$\T$ in terms of $2^n\times2^n$ spin matrices $\X_\nu$, $\Y_\nu$, $\Z_\nu$.
A rescaled transfer matrix $\V$ is defined whose eigenvalues are, up to a
trivial factor, the eigenvalues of $\T$.
We define further spin matrices $\Ga_\nu$ and two commuting projection
classes $\J_{\al\be}^+$ and $\J_{\al\be}^-$ of their bilinears.
After investigating the relevant properties of the $\J_{\al\be}^\pm$,
we express $\V$ in terms of them.
We introduce $2n\times2n$ matrices $J_{\al\be}$ whose algebra is identical
to that of $\J_{\al\be}^\pm$ and define matrices $V^\pm$ in terms of the
$J_{\al\be}$ such that the relation between $V^\pm$ and $J_{\al\be}$ is
closely related to that between $\V$ and $\J_{\al\be}^\pm$.
The result of the well-known diagonalization procedure for the $V^\pm$
is given and the analogy between $V^\pm$ and $\V$ exploited for the
diagonalization of $\V$.
The eigenvalues of $\V$ and the partition function are found explicitly.

The energy is given by $E=E_a+E_b$ with
\beq
E_a=J_a\sum_{\mu=1}^m\sum_{\nu=1}^ns_{\mu\nu}s_{\mu+1,\nu},~~~~
E_b=J_b\sum_{\mu=1}^m\sum_{\nu=1}^ns_{\mu\nu}s_{\mu,\nu+1},
\eeq
with $\be^{-1}=kT$.
$J_a$ and $J_b$ are temperature-independent interaction energy
parameters that we assume to be negative.
We identify rows $1$ and $m{+}1$ and columns $1$ and $n{+}1$, i.e.,
the lattice is wrapped on a torus.
The $s_{\mu\nu}$ can take the values $\pm1$.
The partition function is then given by
\beq
\label{psum}
Z(a,b)=\sum_{s_{11}}\cdots\sum_{s_{mn}}\exp(-\be E),
\eeq
with the definitions $a=-\be J_a$ and $b=-\be J_b$.
$Z$ can be expressed with the help of a $2^n\times2^n$ transfer matrix $\T$,
$Z(a,b)=\mbox{Tr}\,\T^m$ with $\T$ defined by its elements
($s_{n+1}\equiv s_1$),
\beq
\langle\pi|\T|\pi'\rangle
=\prod_{\nu=1}^n\exp(as_\nu s_\nu'+bs_\nu s_{\nu+1}),
\eeq
where $\pi_\mu=\{s_{\mu1},\ldots,s_{\mu n}\}$ for $\mu=1,\ldots,m$.
We can split $\T$ into a product of two matrices $\T=\Vb\Vap$,
defining $\Va'$ and $\Vb$ by their elements 
\bea
\langle\pi|\Vap|\pi'\rangle
&=&\prod_{\nu=1}^n\exp(as_\nu s_\nu'),
\\
\langle\pi|\Vb|\pi'\rangle
&=&\prod_{\nu=1}^n
\de_{s_\nu s_\nu'}\exp(bs_\nu s_{\nu+1}).
\eea

With the help of the Pauli matrices $\si_x$, $\si_y$, $\si_z$, and the
$2\times2$ unit matrix $\unit$, we define Hermitian $2^n\times2^n$ spin
matrices by the direct products
\beq
\X_\nu=
\left(\ba{c}\scs \nu-1\\\bigotimes\\\scs \nu'=1\ea\unit\right)
\otimes\si_x\otimes
\left(\ba{c}\scs n\\\bigotimes\\\scs \nu''=\nu+1\ea\unit\right),
\eeq
and analogously for $Y_\nu$ and $\Z_\nu$ in terms of $\si_y$ and
$\si_z$, respectively.
With $\abar>0$ defined by $\sinh(2\abar)\sinh(2a)=1$, we can write
$\Vap=[2\sinh(2a)]^{n/2}\Va$ with
\beq
\Va=\prod_{\nu=1}^n\exp\left(\abar\X_\nu\right),
\eeq
and
\beq
\label{vb}
\Vb=\prod_{\nu=1}^n\exp\left(b\Z_\nu\Z_{\nu+1}\right),
\eeq
where we have identified $\Z_{n+1}=\Z_1$.
The transfer matrix may then be expressed as $\T=[2\sinh(2a)]^{n/2}\Vb\Va$.
Due to the cyclic property of the trace, we may rewrite the partition
function (\ref{psum}) as
\beq
Z(a,b)=[2\sinh(2a)]^{mn/2}\mbox{Tr}\V^m,
\eeq
where $\V$ is defined by the Hermitian matrix
\beq
\label{v}
\V=\V_{a/2}\Vb\V_{a/2}
\eeq
with
\beq
\label{vahalf}
\V_{a/2}=\prod_{\nu=1}^n\exp(\abar\X_\nu/2)
\eeq
so that $\V_{a/2}^2=\Va$.
If $\La_k$ are the $2^n$ eigenvalues of $\V$, we have
\beq
Z(a,b)=[2\sinh(2a)]^{mn/2}\sum_{k=1}^{2^n}\La_k^m.
\eeq
Our task is therefore to find the eigenvalues of $\V$.

Define the $2n$ matrices ($\nu=1,\ldots,n$)
\bea
\Ga_{2\nu-1}
&=&
\X_1\cdots \X_{\nu-1}\Z_\nu,\\
\Ga_{2\nu}
&=&
\X_1\cdots \X_{\nu-1}\Y_\nu,
\eea
which obey
\beq
\{\Ga_\mu,\Ga_\nu\}=2\de_{\mu\nu}.
\eeq
Define further the matrix
\beq
\label{ux}
\U_\X=\X_1\cdots\X_n=i^n\Ga_1\Ga_2\cdots\Ga_{2n},~~~~~~\U_\X^2=\unit,
\eeq
which anticommutes with every $\Ga_\mu$, $\{\Ga_\mu,\U_\X\}=0$.
Now we can write for the matrices appearing in the exponents of
Eqs.\ (\ref{vb}) and (\ref{vahalf})
\bea
\label{x}
\X_\nu&=&-\frac{i}{2}[\Ga_{2\nu},\Ga_{2\nu-1}],~~~
\nu=1,\ldots,n,\\
\Z_\nu\Z_{\nu+1}&=&-\frac{i}{2}[\Ga_{2\nu+1},\Ga_{2\nu}],~~~
\nu=1,\ldots,n-1,\\
\label{zz2}
\Z_n\Z_1&=&\frac{i}{2}\U_\X[\Ga_1,\Ga_{2n}].
\eea

So far our treatment has been rather similar to Huang's write up
\cite{huang} of Kaufman's approach \cite{kaufman}.
Our subsequent treatment rests on the observation that the formulation
(\ref{v}) of $\V$ with Eqs.\ (\ref{vb}) and (\ref{vahalf}) involves only
the product $\U_\X$ of all $\Ga_\nu$ and bilinears
$\Ga_\al\Ga_\be$, see Eqs.\ (\ref{ux})--(\ref{zz2}).
This will allow us to express $\V$ in terms of the elements of two
commuting algebras of projected bilinears of the $\Ga_\nu$.

With the help of the projectors
\beq
\P^\pm\equiv\frac{1}{2}(\unit\pm\U_\X),
\eeq
define the matrices
\beq
\label{jgaga}
\J_{\al\be}=-\frac{i}{4}[\Ga_\al,\Ga_\be],~~~~~~
\J_{\al\be}^\pm=\P^\pm\J_{\al\be},~~~~~~
\eeq
so that
\beq
\label{jprops}
\J_{\al\be}=\J_{\al\be}^++\J_{\al\be}^-,~~~~~~
\U_\X\J_{\al\be}^\pm=\pm\J_{\al\be}^\pm.
\eeq
Since $\J_{\al\be}^\pm=-\J_{\be\al}^\pm$, there are $n(2n-1)$ such
independent matrices of each kind $\J_{\al\be}^+$ and $\J_{\al\be}^-$.
It is straightforward to show that their algebra decomposes into
two commuting parts, $[\J_{\al\be}^+,\J_{\ga\de}^-]=0$, which obey identical
algebras
\beq
\label{jalgebra}
[\J_{\al\be}^\pm,\J_{\ga\de}^\pm]
=i(\de_{\al\ga}\J_{\be\de}^\pm+\de_{\be\de}\J_{\al\ga}^\pm
-\de_{\al\de}\J_{\be\ga}^\pm-\de_{\be\ga}\J_{\al\de}^\pm).
\eeq

Next note that with Eqs.\ (\ref{x})--(\ref{zz2}), (\ref{jgaga}), and
(\ref{jprops}) we can write
\bea
\X_\nu&\!=\!&2(\J_{2\nu,2\nu-1}^+{+}\J_{2\nu,2\nu-1}^-),~~~
\nu=1,\ldots,n,
\\
\!\!\!\!\Z_\nu\Z_{\nu+1}&\!=\!&2(\J_{2\nu+1,2\nu}^+{+}\J_{2\nu+1,2\nu}^-),~~~
\nu=1,\ldots,n{-}1,
\\
\Z_n\Z_1&\!=\!&-2\U_\X(\J_{1,2n}^+{+}\J_{1,2n}^-)
=-2(\J_{1,2n}^+{-}\J_{1,2n}^-).
\eea
This allows us to express $\V_{a/2}$ from Eq.\ (\ref{vahalf}) and $\Vb$
from Eq.\ (\ref{vb}) in terms of the $\J_{\al\be}^\pm$,
\beq
\V_{a/2}=\prod_{\nu=1}^n\exp[\abar(\J_{2\nu,2\nu-1}^++\J_{2\nu,2\nu-1}^-)]
=\V_{a/2}^+\V_{a/2}^-
\eeq
with
\beq
\label{vapm}
\V_{a/2}^\pm=\prod_{\nu=1}^n\exp(\abar\J_{2\nu,2\nu-1}^\pm),
\eeq
and
\bea
\Vb
&=&\exp[-2b(\J_{1,2n}^+{-}\J_{1,2n}^-)]
\nn\\
&&\times
\prod_{\nu=1}^{n-1}\exp[2b(\J_{2\nu+1,2\nu}^+{+}\J_{2\nu+1,2\nu}^-)]
\nn\\
&=&\Vb^+\Vb^-
\eea
with
\beq
\label{vbpm}
\Vb^\pm=\exp(\mp2b\J_{1,2n}^\pm)
\prod_{\nu=1}^{n-1}\exp(2b\J_{2\nu+1,2\nu}^\pm).
\eeq
The rescaled transfer matrix $\V$ defined in Eq.\ (\ref{v}) reads
then $\V=\V^+\V^-$ with
\beq
\label{tvv}
\V^\pm=\V_{a/2}^\pm\Vb^\pm\V_{a/2}^\pm,~~~~~~[\V^+,\V^-]=0.
\eeq

Define $N\times N$ matrices $J_{\al\be}$ by their elements
\beq
\left(J_{\al\be}\right)_{ij}
=-i(\de_{\al i}\de_{\be j}-\de_{\be i}\de_{\al j}),
\eeq
where Greek and Latin indices run from 1 to $N$.
Since $J_{\al\be}=-J_{\be\al}$, there are $N(N-1)/2$ such independent
matrices.
As can be easily checked, they obey the algebra (\ref{jalgebra}),
if we set $N=2n$.
Now consider the matrices $S=\exp(ic_{\al\be}J_{\al\be})$, where
$c_{\al\be}$ are arbitrary complex numbers.
The matrices $S$ form the group SO$(N,C)$ of complex $N\times N$
matrices with $S^T=S^{-1}$, $\det S=1$.

Define the SO$(2n,C)$ matrices
\beq
V^\pm=V_{a/2}V_b^\pm V_{a/2}
\eeq
with
\beq
V_{a/2}=\prod_{\nu=1}^n\exp(\abar J_{2\nu,2\nu-1})
\eeq
and
\beq
V_b^\pm=\exp(\mp2b J_{1,2n})
\prod_{\nu=1}^{n-1}\exp(2b J_{2\nu+1,2\nu}),
\eeq
in analogy with $\V^\pm$, $\V_{a/2}^\pm$, and $\Vb^\pm$ in Eqs.\ (\ref{tvv}),
(\ref{vapm}), and (\ref{vbpm}).
Since $\abar$ and $b$ are real, the matrices $V_{a/2}$, $V_b^\pm$, and
$V^\pm$ are not only orthogonal, but also Hermitian, so the $V^\pm$ have only
real eigenvalues and in each case a complete set of orthonormal eigenvectors.

It is well known \cite{kaufman,huang} how to diagonalize matrices of the
types $V^\pm$.
Applying similarity transformations
$V_S^\pm\equiv S_\pm V^\pm S_\pm^{-1}$
with certain explicitly known matrices $S_\pm$, one obtains
\beq
\label{vspm}
V_S^\pm=\exp\left(\sum_{\nu=1}^n
\ga_{\cd{$2\nu{-}1$}{$2\nu{-}2$}}J_{2\nu,2\nu-1}\right)
\eeq
with $\ga_k$ defined by
\beq
\label{gak}
\cosh\ga_k=\cosh2\abar\cosh2b-\cos\frac{\pi k}{n}\sinh2\abar\sinh2b.
\eeq
We fix the sign of $\ga_k$ by defining $\ga_k=2\abar$ for $b=0$
and then analytically continuing to other values of $b$.
For $k=1,\ldots,2n-1$, this means $\ga_k>0$.
On the other hand, for $\ga_0$ this means
\beq
\ga_0=2(\abar-b).
\eeq
Our sign convention for the $\ga_k$ and in particular for $\ga_0$
allows us to treat all $\ga_k$ on an equal footing for both $\abar>b$
and $\abar<b$, i.e., irrespective of the temperature.
Note that the $V_S^\pm$ are only block diagonal with $2\times2$ blocks.
It would be trivial to diagonalize $\V_S^\pm$, but the form given by
Eq.\ (\ref{vspm}) is most convenient for our purposes.

It is straightforward to show that the matrices $S_\pm$ are elements of
SO$(N,C)$ and may therefore be written as
$S_\pm=\exp(ic_{\al\be}^\pm J_{\al\be})$.
Now use the same parameters $c_{\al\be}^\pm$ to define the
$2^n\times2^n$-dimensional transformation matrix
\beq
\S=\S_+\S_-,~~~~~~\S_\pm=\exp(ic_{\al\be}^\pm\J_{\al\be}^\pm),
\eeq
and write
\beq
\V_\S=\S\V\S^{-1}=\S_+\V^+\S_+^{-1}\S_-\V^-\S_-^{-1}
\equiv\V_S^+\V_S^-.
\eeq
The factors defining $\V_S^\pm$ in terms of $\abar$, $b$, $c_{\al\be}^\pm$,
and the $\J_{\al\be}^\pm$ have the same structure as the $V_S^\pm$ in terms
$\abar$, $b$, $c_{\al\be}^\pm$, and the $J_{\al\be}$.
Now imagine using the Baker-Campbell-Hausdorff formula \cite{cbh}
\bea
\lefteqn{\exp(A)\exp(B)=}
\nn\\
&&\textstyle
\exp(A{+}B{-}\frac{1}{2}[B,A]{+}\frac{1}{12}\{[A,[A,B]]{+}[B,[B,A]]\}
+\cdots)
\nn\\
\eea
to work out all products of exponentials in $\V_S^\pm$.
Since the $\J_{\al\be}^+$, $\J_{\al\be}^-$, and $J_{\al\be}$ obey identical
algebras, the result is
\beq
\V_\S^\pm=\exp\left(\sum_{\nu=1}^n
\ga_{\cd{$2\nu{-}1$}{$2\nu{-}2$}}\J_{2\nu,2\nu-1}^\pm\right),
\eeq
so that
\bea
\V_\S
&=&
\exp\left[\sum_{\nu=1}^n
(\ga_{2\nu-1}\J_{2\nu,2\nu-1}^++\ga_{2\nu-2}\J_{2\nu,2\nu-1}^-)\right]
\nn\\
&=&
\exp\bigg[\frac{1}{4}\sum_{\nu=1}^n\ga_{2\nu-1}(\unit+\U_\X)\X_\nu
\nn\\
&&~~~~~~
+\frac{1}{4}\sum_{\nu=1}^n\ga_{2\nu-2}(\unit-\U_\X)\X_\nu\bigg],
\eea
with the same $\ga_k$ as defined in Eq.\ (\ref{gak}) and the subsequent
sign convention.

To diagonalize $\V_\S$, define another similarity transformation
$\V_\Y=\R_\Y\V_\S\R_\Y^{-1}$ with $\R_\Y$ and its inverse given by
\beq
\R_\Y^{\pm1}=2^{-n/2}\prod_{\nu=1}^n(\unit\pm i\Y_\nu).
\eeq
Since $\R_\Y\X_\nu\R_\Y^{-1}=\Z_\nu$, this transformation takes $\V_\S$
into
\bea
\lefteqn{\V_\Y
=\R_\Y\V_\S\R_\Y^{-1}}
\nn\\
&\!=&
\exp\left[\frac{1}{4}\sum_{\nu=1}^n\ga_{2\nu-1}(\unit{+}\U_\Z)\Z_\nu
{+}\frac{1}{4}\sum_{\nu=1}^n\ga_{2\nu-2}(\unit{-}\U_\Z)\Z_\nu\right]
\nn\\
\eea
with $\U_\Z=\Z_1\cdots\Z_n$.

The matrix $\V_\Y$ is diagonal, but we still have to determine its elements.
$\U_\Z$ is a diagonal matrix with elements $+1$ and $-1$ occurring in equal
numbers.
For each element holds, if an even (odd) number of $\Z_\nu$ provides a
factor $-1$, the matrix element of $\U_\Z$ is $+1$ ($-1$).
This means (i) a matrix element of $(\unit+\U_\Z)/2$ is $1$ ($0$) if an
even (odd) number of $\Z_\nu$ provides a factor $-1$,
(ii) a matrix element of $(\unit-\U_\Z)/2$ is $1$ ($0$) if an odd (even)
number of $\Z_\nu$ provides a factor $-1$.
It follows that the $2^n$ eigenvalues of $\V$ split into $2^{n-1}$
eigenvalues of the form
\beq
\label{oddevs}
\exp\left(\frac{1}{2}\sum_{\nu=1}^n(\pm)\ga_{2\nu-1}\right),
\eeq
and $2^{n-1}$ eigenvalues of the form
\beq
\label{evenevs}
\exp\left(\frac{1}{2}\sum_{\nu=1}^n(\pm)\ga_{2\nu-2}\right),
\eeq
where in the first (second) case all sign combinations with an even
(odd) number of minus signs occur.
This is reflected by the indices ``e'' and ``o'' in our result for the
partition function,
\begin{widetext}
\bea
\lefteqn{Z(a,b)
=[2\sinh(2a)]^{mn/2}
\left[\sum_{\rm e}\exp\left(\frac{m}{2}
\sum_{\nu=1}^n(\pm)\ga_{2\nu-1}\right)
+\sum_{\rm o}\exp\left(\frac{m}{2}
\sum_{\nu=1}^n(\pm)\ga_{2\nu-2}\right)\right]}
\nn\\
&=&
\frac{1}{2}[2\sinh(2a)]^{mn/2}
\nn\\
&&\times
\left\{
\prod_{k=1}^n\left[2\cosh\left(\frac{m}{2}\ga_{2k-1}\right)\right]
+\prod_{k=1}^n\left[2\sinh\left(\frac{m}{2}\ga_{2k-1}\right)\right]
+\prod_{k=1}^n\left[2\cosh\left(\frac{m}{2}\ga_{2k-2}\right)\right]
-\prod_{k=1}^n\left[2\sinh\left(\frac{m}{2}\ga_{2k-2}\right)\right]
\right\}.
\nn\\
\eea
\end{widetext}
The last term within the braces has a sign differing from that in
\cite{kaufman}.
This is due to our different sign convention for $\ga_0$.
The eigenvalues of $\T$ are of course obtained by multiplying
Eqs.\ (\ref{oddevs}) and (\ref{evenevs}) with the trivial factor
$[2\sinh(2a)]^{n/2}$.

The results for the eigenvalues of $\T$ and the partition function are
the starting point for the analysis of the thermodynamic properties of
the two-dimensional Ising model, the most interesting case being the
thermodynamic limit $m,n\rightarrow\infty$.
Such analyses can now proceed as usual (see, e.g.,
\cite{onsager,kaufman,huang}) and will not be repeated here.

\end{document}